\documentclass[sigconf]{acmart}

\usepackage{makecell}
\usepackage{enumitem}
\usepackage{multirow}
\usepackage{longtable}
\usepackage{graphicx}
\usepackage{caption}
\usepackage{subcaption}
\usepackage{tabularx}
\usepackage{soul}
\usepackage{float}
\usepackage{url}
\definecolor{darkgreen}{rgb}{0.0, 0.5, 0.0}
\definecolor{navyblue}{rgb}{0.0, 0.0, 0.5}

\newif\ifredact
\redactfalse  

\newcommand{\ts}{\textsc{TalkSketchD}}

\newcommand{\dquote}[1]{\textit{``#1''}}

\newif\ifcomment
\commenttrue   
\ifcomment
  \newcommand{\missing}[1]{\textcolor{red}{~#1}}
  \newcommand{\wei}[1]{~\sethlcolor{cyan!40}\hl{[Weiyan: #1]}}
  \newcommand{\ken}[1]{~\sethlcolor{yellow!60}\hl{[Kenny: #1]}}
  \newcommand{\dor}[1]{~\sethlcolor{green!40}\hl{[Dorien: #1]}}
  \newcommand{\rrev}[3]{\textcolor{blue}{[RevID #1] #2: #3}} 
\else
  \newcommand{\missing}[1]{}
  \newcommand{\wei}[1]{}
  \newcommand{\ken}[1]{}
  \newcommand{\dor}[1]{}
  \newcommand{\rrev}[3]{}
\fi

\AtBeginDocument{%
  }

\setcopyright{acmlicensed}
\copyrightyear{2026}
\acmYear{2026}
\setcopyright{cc}
\setcctype{by}
\acmConference[DIS Companion '26]{Designing Interactive Systems Conference}{June 13--17, 2026}{Singapore, Singapore}
\acmBooktitle{Designing Interactive Systems Conference (DIS Companion '26), June 13--17, 2026, Singapore, Singapore}
\acmDOI{10.1145/3802974.3809463}
\acmISBN{979-8-4007-2632-3/2026/06}

\begin{document}

\title[Exploring Spontaneous Speech with Sketch for Intent Alignment in Multimodal LLMs]{When Drawing Is Not Enough: Exploring Spontaneous Speech with Sketch for Intent Alignment in Multimodal LLMs}
\author{Weiyan Shi}
\email{weiyanshi6@gmail.com}
\orcid{0009-0001-6035-9678}
\affiliation{
  \institution{Singapore University of Technology and Design}
  \country{Singapore}\city{Singapore}
  \country{Singapore}
}

\author{Dorien Herremans}
\email{dorien_herremans@sutd.edu.sg}
\orcid{0000-0001-8607-1640}
\affiliation{
  \institution{Singapore University of Technology and Design}
  \country{Singapore}\city{Singapore}
  \country{Singapore}
}

\author{Kenny Tsu Wei Choo}
\email{kennytwchoo@gmail.com}
\orcid{0000-0003-3845-9143}
\affiliation{
  \institution{Singapore University of Technology and Design}
  \city{Singapore}
  \country{Singapore}
}

\ccsdesc[500]{Human-centered computing~Interactive systems and tools}
\ccsdesc[500]{Human-centered computing~Human computer interaction (HCI)}

\keywords{large language model, sketching, multimodal interaction, image generation, speech, human-AI alignment}


\begin{abstract}
Early-stage design ideation often relies on rough sketches created under time pressure, leaving much of the designer’s intent implicit. In practice, designers frequently speak while sketching, verbally articulating functional goals and ideas that are difficult to express visually. 
We introduce \ts{}, a sketch-while-speaking dataset that captures spontaneous speech temporally aligned with freehand sketches during early-stage toaster ideation. To examine the dataset’s value, we conduct a sketch-to-image generation study comparing sketch-only inputs with sketches augmented by concurrent speech transcripts using multimodal large language models (MLLMs). Generated images are evaluated against designers’ self-reported intent using a reasoning MLLM as a judge. 
Quantitative results show that incorporating spontaneous speech significantly improves judged intent alignment of generated design images across form, function, experience, and overall intent. These findings demonstrate that temporally aligned sketch-and-speech data can enhance MLLMs’ ability to interpret user intent in early-stage design ideation.
\end{abstract}

\maketitle

\section{Introduction}

Sketching plays a central role in early-stage design ideation\cite{buxton2010sketching}.
Designers often rely on quick, rough sketches to externalise ideas under time constraints~\cite{suwa2022roles}, prioritising speed and exploration over visual precision.
As a result, sketches produced during ideation are frequently ambiguous and underspecified.
At the same time, designers typically hold richer intentions in mind than what is explicitly captured on the page, including functional goals, usage scenarios, and interaction concepts that are difficult to draw in the moment.

In natural design practice, designers commonly speak while sketching~\cite{kim2009study, purcell1998drawings}. Such spontaneous speech may include fragmented descriptions, explanations, or reflections that accompany the drawing process. Unlike think-aloud protocols~\cite{ericsson2017protocol}, which explicitly instruct designers to continuously externalize their thought process, this speech arises naturally as designers verbally supplement and interpret their sketches — articulating intent that the drawing alone cannot fully convey. Although this speech is often informal and imperfectly articulated, it can provide valuable cues about the designer's underlying intent. However, most sketch-based AI-supported ideation systems rely solely on visual input or carefully constructed textual prompts~\cite{peng2024designprompt, lin2025inkspire, davis2025sketchai, shi2026taxonomy}, overlooking the role of spontaneous speech as part of the ideation process.

Motivated by this gap, we ask:
\textbf{\emph{How do sketching and spontaneous speech complement each other in expressing design intent, and what value does capturing them together offer for AI-supported design ideation?}}
To explore this question, we introduce \ts{}, a sketch-while-speaking dataset collected from an in-lab early-stage design study.
In the dataset, participants freely sketched design concepts while verbally expressing their thoughts, resulting in temporally aligned sketches and spontaneous speech produced during the same ideation process.
Importantly, \ts{} differs from existing sketch datasets that focus on sketch–label or sketch–text alignment, such as \textit{Quick, Draw!}~\cite{ha2017neural} and ShoeV2~\cite{pang2019generalising}. 
In these datasets, textual annotations are typically created post hoc as category labels or curated descriptions of completed sketches. 
In contrast, \ts{} captures spontaneous speech produced concurrently with sketching, providing access to designers’ intent as it unfolds during ideation rather than as a retrospective annotation.

To provide an initial assessment of the dataset’s potential value, we employ multimodal large language models (MLLMs) to generate design images from participants’ sketches. We compare two input conditions—\textit{sketch-only} and \textit{sketch+speech}—while keeping the visual sketch input identical, allowing us to isolate the effect of concurrent speech transcripts on the generated outputs. 
We then evaluate the generated images against designers’ self-reported design intent, using reasoning MLLM as a judge~\cite{chen2024mllm, chen2024humans} to assess intent alignment.

This work makes the following contributions:
\begin{itemize}
    \item We introduce \ts{}, the first dataset capturing designers sketching while speaking during early-stage design ideation, reflecting natural, time-pressured creative practice.
    \item We present an initial exploration of the dataset’s usefulness through a sketch-to-image generation task, comparing sketch-only and sketch--speech inputs and evaluating their alignment with designer intent.
\end{itemize}

\section{Dataset}
\label{sec:dataset}

We introduce \ts{}, a work-in-progress sketch-and-speech dataset collected from an in-lab design study with 11 participants. The dataset comprises 58 sketch-and-speech instances capturing sketch-while-speaking behaviour during AI-assisted design ideation, each paired with its sketch context and a self-reported intent summary.

\textit{\textbf{Participants and task.}}
\ts{} comprises design sessions from $N{=}11$ participants. Each participant completed a 30-minute design task: \dquote{freely ideate toaster concepts during a 30-minute session.} Participants were instructed to explore as many design ideas as possible within the allotted time, without aiming to converge on a single final concept.

\textit{\textbf{Data collection.}}
Data were collected in a controlled laboratory setting using an AI-assisted sketching interface, \textsc{TalkSketch}, which records speech in real time~\cite{shi2026talksketch}. Participants completed an open-ended early-stage product design ideation task, sketching freely while verbally articulating their ideas as they emerged. All sessions were conducted on an iPad Pro 13" with an Apple Pencil Pro, with screen capture used to record the full sketching process. Participants’ speech was recorded concurrently and transcribed in real time using Google Cloud Speech-to-Text\footnote{\url{https://cloud.google.com/speech-to-text}}. This study was reviewed and approved by the Institutional Review Board of Singapore University of Technology and Design (IRB approval no. IRB-25-00717), and all participants provided informed consent.

\begin{figure*}[ht] \centering \includegraphics[width=\linewidth]{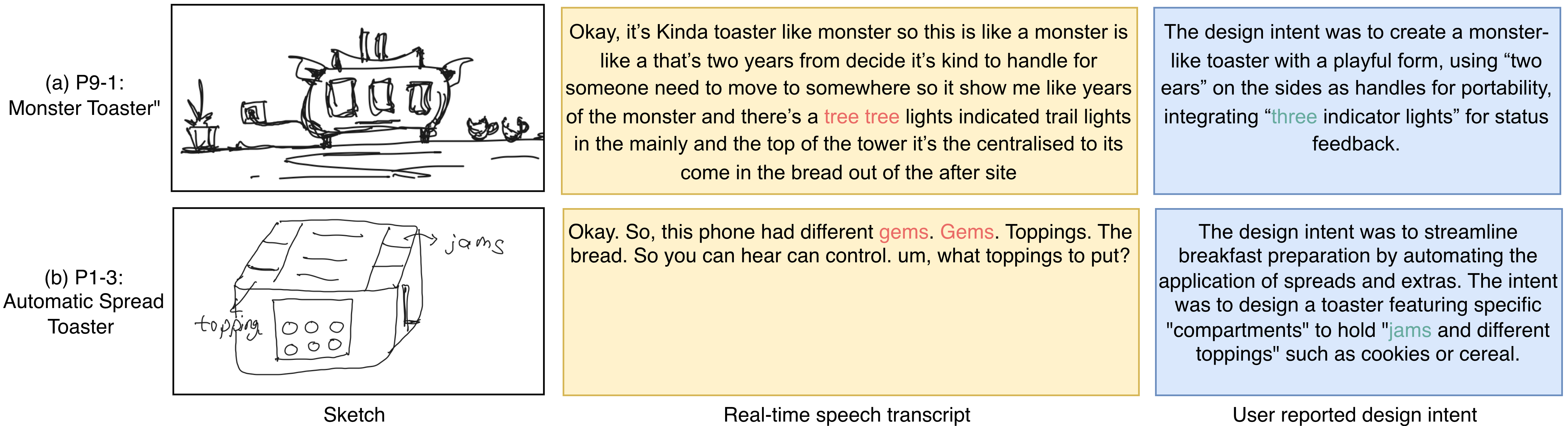} \caption{Two example instances from the dataset, illustrating how early-stage sketches, real-time speech, and post-task intent reports jointly capture design intent.
\textbf{Left:} freehand sketches produced during ideation.
\textbf{Middle:} real-time speech transcripts recorded while sketching, containing recognition errors and disfluencies typical of spontaneous speech (e.g., \dquote{gems} instead of \dquote{jams} and \dquote{tree} instead of \dquote{three}, repeated or fragmented phrases).
\textbf{Right:} participants’ self-reported design intent collected after the session, which consolidates and clarifies the intended concept.
The top example (\textbf{Participant~9, Instance~1, Monster Toaster}) presents a function-driven concept that automates the application of spreads and toppings via dedicated compartments, while the bottom example (\textbf{Participant~1, Instance~3, Automatic Spread Toaster}) depicts a playful, form-driven monster-like toaster with expressive features and three visual status indicators.}
\label{fig:dataset-example}
\end{figure*}

\textbf{\textit{Dataset structure.}}
\ts{} is organised as a collection of multimodal sketch-and-speech instances. Each instance (Figure~\ref{fig:dataset-example}) comprises:
(1) continuous freehand sketch data captured from a digital canvas,
(2) time-aligned speech recordings and their corresponding real-time transcripts produced during sketching, and
(3) brief self-reported descriptions summarising the intended design concepts, which serve as high-level representations of design intent.

We intentionally retain the original real-time transcripts without manual correction, preserving transcription errors to reflect realistic noisy speech conditions encountered in practical AI-assisted ideation settings.

Figure~\ref{fig:dataset-example} illustrates two common patterns observed in the dataset.
In some cases (e.g., Figure~\ref{fig:dataset-example} (a), Participant~9, Instance~1), designers extensively verbalised ideas while sketching, providing rich supplementary information.
In contrast, other cases (e.g., Figure~\ref{fig:dataset-example} (b), Participant~1, Instance~3) involved relatively limited verbalisation during sketching, with key aspects of functional intent articulated more clearly only in the post-task self-report.
Across both patterns, a common challenge is the pervasive presence of noise in the speech transcripts, including recognition errors and disfluencies.

This observation also reveals a key challenge in dataset construction: spontaneous speech that lacks rich semantic content offers limited added value as a supplementary modality, reducing the effectiveness of language as an extension of sketch-based intent representation.

\section{Method}
\label{sec:method}

To probe the potential value of the \ts{}, we designed an initial exploratory task using sketch-based design image generation. The goal of this experiment is to examine whether spontaneous speech captured during sketching provides complementary semantic information beyond sketches alone, and whether this additional information influences downstream generative outcomes.

Sketch-based design image generation was selected as a representative task because it aligns with how designers commonly engage with AI during early-stage ideation.
Rather than crafting detailed prompts, designers often use sketch-to-image tools to externalise and iterate on emerging ideas.
Within this workflow, sketches express visual structure, while spoken descriptions naturally accompany drawing to articulate intent that is not yet visually specified.
Real-time speech transcripts are therefore imperfect by nature, capturing the realities of design interaction rather than idealised input.

As an initial study, this experiment aims to quantitatively characterise differences between \textit{Sketch-only} and \textit{Sketch+Speech} conditions using a sketch-to-image generation task.
We complement the quantitative analysis with a qualitative case to help interpret how spontaneous speech contributes to intent-aligned generation in specific instances.
This analysis is intended to demonstrate the applied relevance and analytic potential of the sketch-and-speaking dataset.

\textit{\textbf{Sketch-to-image generation.}}
To examine how speech transcripts influence sketch-based design image generation, we conducted image generation under two controlled input conditions. In both conditions, the same freehand sketch was provided to the image generation model, with the only difference being whether user speech was included.

In the \textit{Sketch-only} condition, the model was prompted as: \dquote{Render this as a real toaster based on the sketch.}
In the \textit{Sketch+Speech} condition, the same sketch was provided together with the corresponding real-time speech transcript captured during sketching. The model was prompted as: \dquote{Render this as a real toaster based on the sketch and user prompt.}
In both conditions, speech transcripts were included verbatim without manual correction, preserving recognition errors typical of real-time transcription. All images were generated using Gemini~2.5~Flash~Image (Nano Banana)\footnote{https://docs.cloud.google.com/vertex-ai/generative-ai/docs/models/gemini/2-5-flash-image}.

\textit{\textbf{MLLM-as-judge for intent alignment.}}
To compare which generated image better reflected the user’s intended design concept, we adopted an MLLM-as-judge evaluation approach~\cite{chen2024mllm, chen2024humans}. A multimodal reasoning model (Gemini~2.5~Pro\footnote{https://docs.cloud.google.com/vertex-ai/generative-ai/docs/models/gemini/2-5-pro}) was configured to act as a design professional and to evaluate two generated toaster designs for each design instance.

For each comparison, the judge was provided with the participant’s self-reported design intent together with two generated images (Image~A and Image~B). The judge was prompted as follows:

\begin{quote}\ttfamily
\small
You are a design professional. Evaluate two toaster designs (A and B) against a user's real intent. \textbf{Criteria (Score 1--7):} (1) Form: Does the visual appearance (shape, color, material, style) match the intent? (2) Function: Does the design include the specific features and components described? (3) Experience: Does the design reflect the intended interaction experience, usage flow, and conceptual goal described by the user? \textbf{Task:} Rate Image~A and Image~B on each criterion, provide an overall intent alignment score (1--7) for both, and identify a winner.
\end{quote}

The judge produced structured scores for each criterion, an overall intent alignment score, a binary judgement, and a short qualitative rationale. This evaluation focuses on relative Human-AI alignment between conditions.
\section{Results}
\label{sec:results}

Using \ts{}, we generated a total of 116 design images, corresponding to 58 design instances under two conditions (sketch-only and sketch-plus-speech). For each instance, the same sketch was used to produce a paired set of images, resulting in one image per condition. Each image pair was evaluated using the MLLM-as-judge protocol. For every pair, the judge provided scores along three dimensions—form, function, and experience—as well as an overall intent alignment score. In addition, the judge identified a winner for each pair, indicating which image was better aligned with the participant’s reported design intent.

\begin{figure*}[h]
    \centering
    \includegraphics[width=0.8\linewidth]{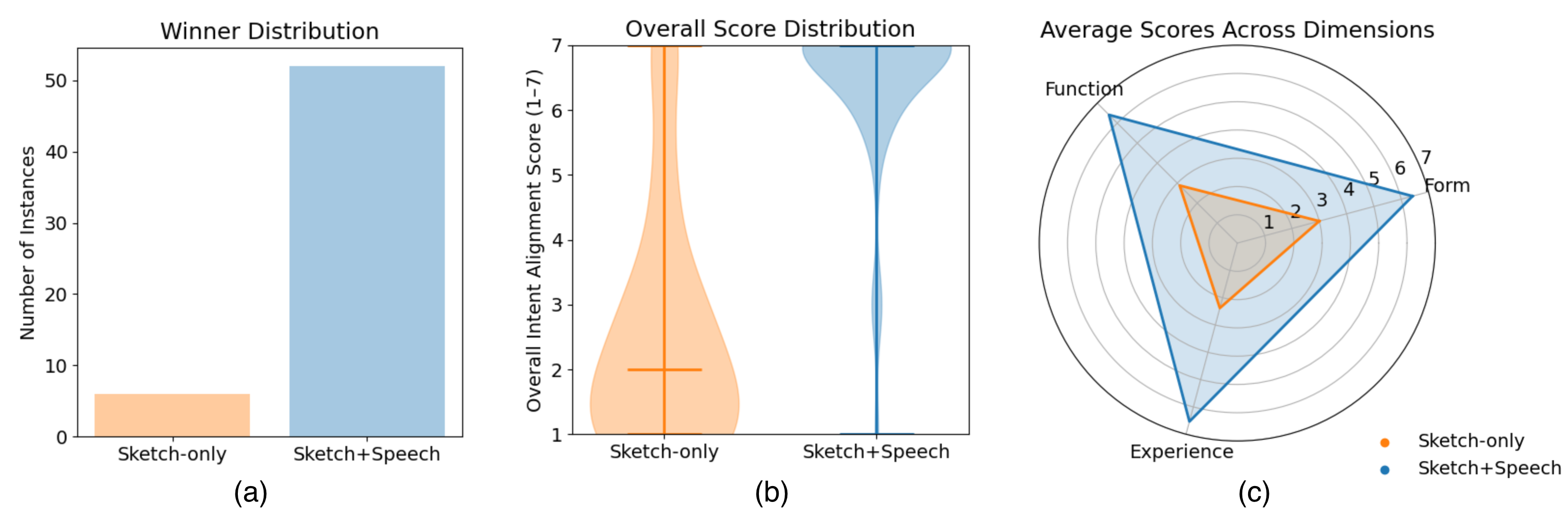}
    \caption{
    Summary of MLLM-as-judge evaluations comparing Sketch-only and Sketch--Speech conditions across the \ts{} instances.
    \textbf{Left: Winner distribution} shows the number of instances where each condition was judged to better align with the reported intent.
    \textbf{Middle: Overall score distribution} visualises the distribution of overall intent alignment scores (1--7) using violin plots with median, highlighting central tendency and variability across instances.
    \textbf{Right: Average scores across dimensions} reports the mean scores for \emph{Form}, \emph{Function}, and \emph{Experience}, providing a dimension-level view of intent alignment.}
    \label{fig:judge-results}
\end{figure*}

\textit{\textbf{Quantitative Results.}}
Across 58 design instances, clear differences emerged between the two conditions (Figure~\ref{fig:judge-results}).
For all four evaluation metrics—form, function, experience, and overall intent alignment—the \textit{Sketch-only} condition consistently received low median scores
($Md = 2.0$ across all metrics; $IQR_{\text{form}} = [1.25, 4.0]$, 
$IQR_{\text{function}} = [1.0, 4.0]$, 
$IQR_{\text{experience}} = [1.0, 3.0]$, 
$IQR_{\text{overall}} = [1.0, 3.0]$),
indicating limited alignment with participants’ stated design intent.

In contrast, the \textit{Sketch+Speech} condition achieved substantially higher median scores across all metrics
($Md = 7.0$ for form, function, experience, and overall intent alignment), 
with narrower dispersion
($IQR_{\text{form}} = [6.0, 7.0]$, 
$IQR_{\text{function}} = [6.0, 7.0]$, 
$IQR_{\text{experience}} = [7.0, 7.0]$, 
$IQR_{\text{overall}} = [7.0, 7.0]$).

Mann--Whitney U tests revealed statistically significant differences between conditions for all metrics 
($U_{\text{form}} = 350.5$, 
$U_{\text{function}} = 323.5$, 
$U_{\text{experience}} = 217.0$, 
$U_{\text{overall}} = 265.0$; all $p < 10^{-14}$), 
indicating a substantial improvement in judged intent alignment when spontaneous speech transcripts were incorporated alongside sketches.

\begin{figure*}[h]
    \centering
    \includegraphics[width=\linewidth]{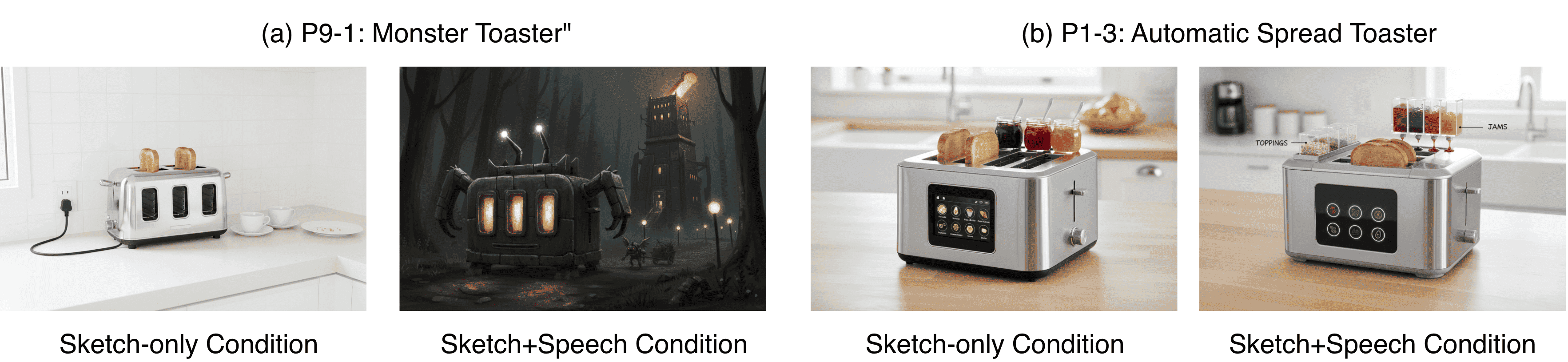}
    \caption{Qualitative comparison of sketch-based image generation under sketch-only and sketch-plus-speech conditions for two design instances.
    \textbf{(a) P9--1 Monster Toaster.}
    Under the sketch-only condition, the generated image depicts a conventional toaster with three front-facing windows, capturing limited aspects of the intended playful character.
    When combined with spontaneous speech, the generation shifts towards a monster-like toaster with expressive features and three central lights; however, erroneous transcript (\dquote{tree} instead of \dquote{three}) content also introduces unintended tree-related environmental context.
    \textbf{(b) P1--3 Automatic Spread Toaster.}
    Both sketch-only and sketch-plus-speech conditions produce a toaster with a touchscreen interface.
    With sketch input alone, topping-related elements appear as externally placed jars, whereas the sketch-plus-speech condition yields a more structurally integrated design with embedded jam and topping compartments, better reflecting the participant’s functional intent.}
    \label{fig:qualitative-case}
\end{figure*}

\textit{\textbf{Two Qualitative Cases.}}
Figure~\ref{fig:qualitative-case} presents two representative design instances comparing sketch-only and sketch-plus-speech image generation (additional examples are shown in Figure~\ref{fig:dataset-example}). Overall, both conditions are capable of producing plausible and visually coherent design outcomes; however, the role of spontaneous speech varies across cases.

In the \emph{Automatic Spread Toaster} example (P1--3), the participant’s reported intent emphasised automating the application of spreads and toppings. While the spoken description itself contained limited additional detail beyond what was visible in the sketch, the sketch-plus-speech generation nevertheless produced a design with more structurally integrated topping compartments, whereas the sketch-only condition depicted toppings as externally placed elements.

At the same time, the \emph{Monster Toaster} example (P9--1) illustrates a complementary limitation: although speech helped steer the generation towards a more expressive, character-like design, transcription errors (e.g., \dquote{tree} instead of \dquote{three}) also introduced unintended \dquote{tree-like} environmental elements.
\section{Conclusion and Discussion}

We introduce \ts{}, a sketch-while-speaking dataset collected from toaster design ideation, and demonstrate that incorporating spontaneous speech significantly improves user intent alignment in sketch-to-image generation.

From a human--computer interaction perspective, this empirical finding provides grounding for treating sketching while speaking as a viable interaction paradigm for AI-supported ideation. 
It suggests that MLLMs are sufficiently capable of leveraging loosely structured, real-time speech to enhance alignment with user intent, opening possibilities for integrating speech more naturally into creative workflows.

From an AI perspective, our results also expose fundamental challenges. 
Spontaneous speech differs from typed prompts in being incremental, difficult to revise, and prone to transcription noise—especially when produced concurrently with sketching. 
The observed performance gains raise important questions about robustness, signal fusion, and interpretability: how multimodal models balance heterogeneous inputs of uneven reliability, how speech and sketch are internally represented and integrated, and how future training and evaluation strategies can better reflect real-world human input.

Beyond immediate performance gains, \ts{} provides an empirical basis for studying alignment under realistic human expression conditions. 
Unlike conventional prompt datasets that assume clean, fully specified instructions, sketch-while-speaking data capture intent as it unfolds—partial, evolving, and embedded in multimodal context. 
This shifts alignment from matching static commands to tracking temporally evolving intent, and invites rethinking evaluation benchmarks to account for process-level interaction. 
Embracing such spontaneous, multimodal input moves alignment research closer to real-world settings, where human communication is inherently dynamic rather than perfectly articulated.

\begin{acks}
This research is supported by the Ministry of Education, Singapore, under its MOE AcRF Tier 1, and funded through the SUTD Kickstarter Initiative (SKI) administered by SUTD (SKI 2021\_05\_16).
Any opinions, findings and conclusions or recommendations expressed in this material are those of the author(s) and do not reflect the views of the Ministry of Education, Singapore.
\end{acks}

\bibliographystyle{ACM-Reference-Format}
\bibliography{main}

@InProceedings{peng2024designprompt,
  author     = {Peng, Xiaohan and Koch, Janin and Mackay, Wendy E},
  booktitle  = {Proceedings of the 2024 ACM Designing Interactive Systems Conference},
  date       = {2024-07},
  title      = {Designprompt: Using multimodal interaction for design exploration with generative ai},
  address    = {New York, NY, USA},
  doi        = {10.1145/3643834.3661588},
  pages      = {804--818},
  publisher  = {ACM},
  series     = {DIS ’24},
  collection = {DIS ’24},
  year       = {2024},
}

@InProceedings{lin2025inkspire,
  author     = {Lin, David Chuan-En and Kang, Hyeonsu B and Martelaro, Nikolas and Kittur, Aniket and Chen, Yan-Ying and Hong, Matthew K},
  booktitle  = {Proceedings of the 2025 CHI Conference on Human Factors in Computing Systems},
  date       = {2025-04},
  title      = {Inkspire: supporting design exploration with generative ai through analogical sketching},
  address    = {New York, NY, USA},
  doi        = {10.1145/3706598.3713397},
  pages      = {1--18},
  publisher  = {ACM},
  series     = {CHI ’25},
  collection = {CHI ’25},
  year       = {2025},
}

@InProceedings{davis2025sketchai,
  author     = {Davis, Richard Lee and Mwaita, Kevin Fred and M{\"u}ller, Livia and Tozadore, Daniel C and Novikova, Aleksandra and K{\"a}ser, Tanja and Wambsganss, Thiemo},
  booktitle  = {Proceedings of the Extended Abstracts of the CHI Conference on Human Factors in Computing Systems},
  date       = {2025-04},
  title      = {SketchAI: A" Sketch-First" Approach to Incorporating Generative AI into Fashion Design},
  address =    {New York, NY, USA},
  doi        = {10.1145/3706599.3719782},
  pages      = {1--7},
  publisher  = {ACM},
  series     = {CHI EA ’25},
  collection = {CHI EA ’25},
  year       = {2025},
}

@InBook{suwa2022roles,
  author       = {Suwa, Masaki and Gero, John S and Purcell, Terry A},
  booktitle    = {Proceedings of the Twentieth Annual Conference of the Cognitive Science Society},
  date         = {2022-05},
  title        = {The Roles of Sketches in Early Conceptual Design Processes},
  address      = {Oxfordshire},
  doi          = {10.4324/9781315782416-188},
  isbn         = {9781315782416},
  pages        = {1043--1048},
  publisher    = {Routledge},
  organization = {Routledge},
  year         = {2022},
}

@Book{buxton2010sketching,
  author    = {Buxton, Bill},
  date      = {2007},
  title     = {Sketching user experiences: getting the design right and the right design},
  address   = {San Francisco, CA, USA},
  doi       = {10.1016/b978-0-12-374037-3.x5043-3},
  isbn      = {9780123740373},
  publisher = {Elsevier},
  year      = {2010},
}

@InProceedings{kim2009study,
  author     = {Kim, Jieun and Bouchard, Carole and Omhover, Jean-Fran{\c{c}}ois and Aoussat, Am{\'e}ziane and Moscardini, Laurence and Chevalier, Aline and Tijus, Charles and Buron, Fran{\c{c}}ois},
  title      = {A study on designer’s mental process of information categorization in the early stages of design},
  booktitle  = {Proceedings of the International Association of Societies of Design Research (IASDR) Conference},
  year       = {2009},
  address    = {Seoul, South Korea},
  publisher  = {The International Association of Societies of Design Research (IASDR)},
  pages      = {2401--2410},
}

@Article{purcell1998drawings,
  author       = {Purcell, A.T. and Gero, J.S.},
  date         = {1998-10},
  journaltitle = {Design Studies},
  title        = {Drawings and the design process: A review of protocol studies in design and other disciplines and related research in cognitive psychology},
  doi          = {10.1016/s0142-694x(98)00015-5},
  issn         = {0142-694X},
  number       = {4},
  pages        = {389--430},
  volume       = {19},
  journal      = {Design studies},
  publisher    = {Elsevier BV},
  year         = {1998},
}

@inproceedings{shi2026talksketch,
  title={TalkSketch: Multimodal Generative AI for Real-Time Sketch Ideation with Speech},
  author={Shi, Weiyan and Upadhyay, Sunaya and Quek, Geraldine and Choo, Kenny Tsu Wei},
  booktitle={International Workshop on Creative AI for Live Interactive Performances},
  pages={83--97},
  year={2026},
  organization={Springer}
}

@article{shi2026taxonomy,
  title={A Taxonomy of Human--MLLM Interaction in Early-Stage Sketch-Based Design Ideation},
  author={Shi, Weiayn and Choo, Kenny Tsu Wei},
  journal={arXiv preprint arXiv:2602.22171},
  year={2026}
}

@inproceedings{chen2024humans,
  title={Humans or LLMs as the Judge? A Study on Judgement Bias},
  author={Chen, Guiming and Chen, Shunian and Liu, Ziche and Jiang, Feng and Wang, Benyou},
  booktitle={Proceedings of the 2024 Conference on Empirical Methods in Natural Language Processing},
  pages={8301--8327},
  year={2024}
}

@article{ha2017neural,
  title={A neural representation of sketch drawings},
  author={Ha, David and Eck, Douglas},
  journal={arXiv preprint arXiv:1704.03477},
  year={2017}
}

@inproceedings{pang2019generalising,
  title={Generalising fine-grained sketch-based image retrieval},
  author={Pang, Kaiyue and Li, Ke and Yang, Yongxin and Zhang, Honggang and Hospedales, Timothy M and Xiang, Tao and Song, Yi-Zhe},
  booktitle={Proceedings of the IEEE/CVF conference on computer vision and pattern recognition},
  pages={677--686},
  year={2019}
}

@article{ericsson2017protocol,
  title={Protocol analysis},
  author={Ericsson, K Anders},
  journal={A companion to cognitive science},
  pages={425--432},
  year={2017},
  publisher={Wiley Online Library}
}

@inproceedings{chen2024mllm,
  title={Mllm-as-a-judge: Assessing multimodal llm-as-a-judge with vision-language benchmark},
  author={Chen, Dongping and Chen, Ruoxi and Zhang, Shilin and Wang, Yaochen and Liu, Yinuo and Zhou, Huichi and Zhang, Qihui and Wan, Yao and Zhou, Pan and Sun, Lichao},
  booktitle={Forty-first International Conference on Machine Learning},
  year={2024}
}

\appendix
\end{document}